\documentclass[showpacs,10pt,aps,pra,notitlepage]{revtex4-1}
\usepackage{hyperref}
\usepackage{amsmath}
\usepackage{amsfonts}
\usepackage{amssymb}
\usepackage{bm}
\usepackage{tensor}
\usepackage{graphicx}
\usepackage[]{subfigure}
\begin{document}

\title{General Relativistic Contributions in Transformation Optics}
\author{Robert T. Thompson}
\email{robert@cosmos.phy.tufts.edu}
\affiliation{Department of Mathematics and Statistics,
University of Otago,
P.O. Box 56, 
Dunedin, 9054,  New Zealand
}

\begin{abstract}
One potentially realistic specification for devices designed with transformation optics is that they operate with high precision in curved space-time, such as Earth orbit.  
This raises the question of what, if any, role does space-time curvature play in determining transformation media?  
Transformation optics has been based on a three-vector representation of Maxwell's equations in flat Minkowski space-time.  
I discuss a completely covariant, manifestly four-dimensional approach that enables transformations in arbitrary space-times, and demonstrate this approach for stable circular orbits in the spherically symmetric Schwarzschild geometry.  Finally, I estimate the magnitude of curvature induced contributions to satellite-borne transformation media in Earth orbit and comment on the level of precision required for metamaterial fabrication before such contributions become important.

\end{abstract}

\pacs{42.15.Eq, 42.70.-a, 78.67.Pt, 81.05.Xj, 91.10.Sp}
\maketitle

\section{Introduction}
Since its recent inception, metamaterial-based transformation optics has shown great promise as a new approach for manipulating electromagnetic fields and designing novel optical and electromagnetic devices, such as the electromagnetic cloak \cite{Pendry2006sc,Leonhardt2006sc,Cummer2006pre,Schurig2006sc,Rahm2007pn}, diffractionless ``superlens'' \cite{Pendry2000prl,Grbic2004prl}, and temporal cloaking \cite{McCall2011jo,Fridman2011x}; for a review, see \cite{cai2009,Leonhardt2009po}.  
The original approach to transformation optics used purely spatial transformations \cite{Pendry2006sc,Schurig2006oe,Milton2006njp}.
A step toward generalizing transformation optics came from matching the constitutive relations for electromagnetic fields propagating in curved vacuum space-times with the constitutive relations for fields in stationary materials residing in Minkowski space-time \cite{Leonhardt2006njp1,Plebanski1960pr,DeFelice1971gerg}.

Common to these initial approaches is that they are based on a 3-vector representation of Maxwell's equations in flat, Minkowski space-time.  
This excludes high precision applications that must take account of general relativistic curvature effects.  Suppose, for example, that one would like to build an orbiting telescope using a superlens based on a space-folding transformation \cite{Leonhardt2006njp1}.  
Would the well-known transformation media required for such a lens in flat space work equally well in orbit or does it require some modification to perform the same function?  
Another example might be a very sensitive satellite antenna designed with transformation optics \cite{Lier2011nat}. 
The 3-vector approach to transformation optics cannot account for space-time effects, but a further generalization of transformation optics has recently been developed that extends the limits of previous approaches \cite{Thompson2011jo1,Thompson2011jo2}.  
This completely covariant, manifestly four-dimensional theory has already been shown to encompass all transformations in a unified framework that can also accommodate relative motion \cite{Thompson2011jo2} and arbitrary, non-vacuum, initial linear dielectric embeddings \cite{Thompson2010pra}.  
Here I demonstrate that this approach also allows for transformations in curved space-times, thereby accounting for possible general relativistic contributions to the resulting transformation media.
  
By using a curved space-time metric instead of the Minkowski metric it is possible to obtain an exact result, but because the scenario under consideration (Earth orbit) is decidedly non-relativistic it is advantageous to look at the non-relativistic limit.  
In particular, we could ask the question: If we design a device in flat (or ``Newtonian'') space-time, how different does the transformation medium need to be for the device to actually work as desired in Earth orbit?  
Given the weakness of gravity, we should expect the answer to be that the necessary material parameters are precisely those expected from the flat space-time calculations, plus something small, plus some higher order terms.

To be more concrete, the envisaged scenario is that of a device in stable circular orbit around a spherically symmetric massive object whose space-time can be described by the Schwarzschild metric.  
We will restrict our attention to radially ingoing waves, and consider transformations over some small solid angle.  
This last restriction ensures that the device can be described in a single locally flat patch by a co-moving observer.  

There is always a choice of coordinate system when specifying optical transformations.  
But in transformation optics, where one wants to specify a transformation that translates into a useful device, some coordinates are more appropriate than others, such as cylindrical coordinates for a cylindrical cloak \cite{Rahm2007pn}.  
This will of course depend on the application in mind, but for the manipulation of radially ingoing light near a massive body, Schwarzschild coordinates are the best choice.  
If one wanted to talk about transformations in the local frame of the device for fields known in the local frame of the device then everything is local and flat, but does not describe the situation under consideration here.  
One could express the incoming fields in the local frame and then talk about everything locally as just mentioned.  
That approach is valid, but still requires all the same machinery developed here, and has the disadvantage of losing the intuitive description of optical transformations that make transformation optics so useful.

There are some complications involved that make identifying and isolating the general relativistic effects more difficult.  
First is the question of how to calculate the purely Newtonian results with which to compare the full results.  
The purely Newtonian results would be those found for a device orbiting around a massive spherically symmetric body in flat space-time, i.e.\ in a theory of gravity where the space-time is not curved by the massive body.  
In flat space-time, the transformation medium for a device moving in rectilinear motion would be obtained through a Lorentz boost \cite{Thompson2011jo2}.  
However, the orbiting device is non-inertial, being accelerated through its orbit, and the full orbital motion cannot be described by a boost.
The best we can do is boost to an instantaneously co-moving frame at some point on the orbit.  
Thus, Sec.\ \ref{Sec:BoostCompare} compares the expanded full results with the instantaneously co-moving device obtained via a boost in flat space-time.

Secondly, the Newtonian limit of the material parameters will still depend on the Newtonian velocity of the device, which for a circular orbit at radius $r$ around a spherically symmetric mass $M$ is $\sqrt{M/r}$.  
To clarify this point, consider for the moment a cloak moving with velocity $v$ through a laboratory in flat space-time.  
The fields measured in the lab frame are not the same as the fields seen by the moving cloak. 
So one must either express the fields in the cloak's frame (via a boost) and do the transformation there, or do all the calculations in the lab frame and then boost the resulting material parameters to the cloak's frame.  
Expanding the final material parameters (in the cloak's frame) in the velocity $v$ finds the same cloak parameters for a stationary cloak, plus some terms that are linear in $v$, plus some higher order terms \cite{Thompson2011jo2}.  
In this example, the term linear in $v$ could be interpreted as a velocity-induced correction to the stationary cloak.  
Returning to the scenario of an orbiting device, velocity dependent terms make up part of the Newtonian result, and care must be taken to isolate the curvature contributions.  
It should be emphasized that because the orbital velocity is expressed in terms of $M$, not every occurrence of $M$ in the expanded material parameters is a curvature contribution.

This article is organized as follows.  
Section \ref{Sec:EandM} briefly summarizes the covariant, four-dimensional theory of electrodynamics.  
Section \ref{Sec:TO} summarizes the covariant approach to transformation optics developed in Refs.\ \cite{Thompson2011jo1,Thompson2011jo2}, and presents the main result found therein as Eq.\ (\ref{Eq:MaterialChi}).  
Section \ref{Sec:GenRelCorr} demonstrates how this approach to transformation optics may be used in a curved space-time; in particular, for a stable circular orbit of the Schwarzschild space-time.  
First I discuss how to express the permeability, permittivity, and magnetoelectric couplings in a local frame co-moving with the material, where these parameters take on their usual physical interpretation.  
This is then demonstrated with some representative transformations.
Section \ref{Sec:Discussion} addresses the question of how these results are different from those in flat space-time and shows that the results cannot be due solely to the orbital motion.  In the non-relativistic, Newtonian limit, the results agree with those of a boost in flat space-time, as expected.  Higher order terms of the expansion contribute curvature induced corrections. I conclude with Sec.\ \ref{Sec:Conclusions}.  I use the geometrized units where $G=c=1$.

\section{Classical Electrodynamics} \label{Sec:EandM}
For details of covariant electromagnetic theory one may consult a number of excellent sources such as \cite{MTW,Post,Lindell,Hehl}, here I describe the salient features of the theory required for our purpose.
Assume space-time to consist of a manifold $M$ and metric $\mathbf{g}$. 
The electric field $\vec{E}$ and magnetic flux $\vec{B}$ are combined into a single mathematical object, the covariant field strength tensor $\mathbf{F}$, that in a local Cartesian frame or Minkowski space-time has the matrix representation
\begin{equation} \label{Eq:FComponents}
 F_{\mu\nu} = \left(
 \begin{matrix}
  0 & -E_x & -E_y & -E_z\\
  E_x & 0 & B_z & -B_y\\
  E_y & -B_z & 0 & B_x\\
  E_z & B_y & -B_x & 0
 \end{matrix}
 \right).
\end{equation}
Additionally, the electric flux $\vec{D}$ and magnetic field $\vec{H}$ are combined into the covariant excitation tensor $\mathbf{G}$, that in a local Cartesian frame or Minkowski space-time has the matrix representation
\begin{equation} \label{Eq:GComponents}
 G_{\mu\nu} = \left(
 \begin{matrix}
  0 & H_x & H_y & H_z\\
  -H_x & 0 & D_z & -D_y\\
  -H_y & -D_z & 0 & D_x\\
  -H_z & D_y & -D_x & 0
 \end{matrix}
 \right).
\end{equation}
Maxwell's equations are succinctly expressed as
$\mathrm{d}\mathbf{F}=0$, and $\mathrm{d}\mathbf{G}=\mathbf{J}$,
where $\mathrm{d}$ is the exterior derivative, and $\mathbf{J}$ is the charge-current 3-form \cite{MTW}.  

In a linear dielectric medium there exists a relationship between $\mathbf{F}$ and $\mathbf{G}$ given by a constitutive relation that may be expressed as \cite{Thompson2011jo1,Thompson2011jo2}
\begin{equation} \label{Eq:Constitutive}
 \mathbf{G} = \boldsymbol{\chi}(\star\mathbf{F}),
\end{equation}
which in component form reads
\begin{equation} \label{Eq:ConstitutiveIndices}
 G_{\mu\nu} = \chi\indices{_{\mu\nu}^{\alpha\beta}}\star\indices{_{\alpha\beta}^{\sigma\rho}}F_{\sigma\rho}.
\end{equation}
In Eq.\ (\ref{Eq:Constitutive}), $\star$ is the Hodge dual on $(M,\mathbf{g})$, which for present purposes is to be understood as a map from 2-forms to 2-forms having components
\begin{equation} \label{Eq:star}
 \star\indices{_{\alpha\beta}^{\mu\nu}} = \frac12 \sqrt{|g|}\epsilon_{\alpha\beta\sigma\rho}g^{\sigma\mu}g^{\rho\nu},
\end{equation}
where $\epsilon_{\alpha\beta\sigma\rho}$ is the totally antisymmetric symbol, and $g$ is the determinant of the metric tensor $g_{\mu\nu}$.  
Also note that $g^{\mu\nu}$ is the matrix inverse of $g_{\mu\nu}$.
The tensor $\boldsymbol{\chi}$ contains information on the dielectric material's properties such as permittivity, permeability, and magnetoelectric couplings.  
We take $\boldsymbol{\chi}$ to be independently antisymmetric on its first two and last two indices, and in vacuum $\boldsymbol{\chi}(\star\mathbf{F}) = \star\mathbf{F}$.  
This last condition means that the classical vacuum is treated as a linear dielectric with trivial $\boldsymbol{\chi}$, recovering the usual, trivial, constitutive relations in vacuum.

The components of the constitutive equation provide a set of six independent equations that in a local frame or Minkowski space-time can be collected in the form
\begin{equation} \label{Eq:ConstitutiveComponents1}
 H_a=(\check{\mu}^{-1})\indices{_a^b}B_b + (\check{\gamma_1}^*)\indices{_a^b}E_b, \  D_a=(\check{\varepsilon}^*)\indices{_a^b}E_b+ (\check{\gamma_2}^*)\indices{_a^b}B_b,
\end{equation}
where the notation $\check{a}$ denotes a $3\times 3$ matrix.  Rearranging these to 
\begin{equation} \label{Eq:ConstitutiveComponents2}
 B_a=(\check{\mu})\indices{_a^b}H_b + (\check{\gamma_1})\indices{_a^b}E_b, \  D_a=(\check{\varepsilon})\indices{_a^b}E_b+ (\check{\gamma_2})\indices{_a^b}H_b,
\end{equation}
gives a representation that may be more familiar and in which subsequent results will be expressed.  
These three-dimensional representations of the completely covariant Eq.\ (\ref{Eq:Constitutive}) are essentially equivalent, and are related by
\begin{equation} \label{Eq:ConstitutiveShift}
 \check{\varepsilon}=\check{\varepsilon}^*-\check{\gamma_2}^*\check{\mu}\check{\gamma_1}^*, \  \check{\gamma_1}=\text{-}\check{\mu}\check{\gamma_1}^*, \  \check{\gamma_2} = \check{\gamma_2}^*\check{\mu}.
\end{equation}
However, one should be aware that these $3\times 3$ matrices are not strictly tensors but simply components of $\boldsymbol{\chi}$ that have been collected into matrices.  The components of $\bm{\chi}$ have been identified elsewhere \cite{Thompson2011jo1,Thompson2011jo2}. 

\section{Transformation Optics} \label{Sec:TO}
To understand transformation optics, start with an initial space-time manifold $(M,\mathbf{g},\star)$, field configuration $(\mathbf{F},\mathbf{G},\mathbf{J})$, and material distribution $\boldsymbol{\chi}$, where $\mathrm{d}\mathbf{F}=0$, $\mathrm{d}\mathbf{G}=\mathbf{J}$, and $\mathbf{G}=\boldsymbol{\chi}(\star\mathbf{F})$.  
Imagine now a map $T:M\to \tilde{M}\subseteq M$ that maps $M$ to some image $\tilde{M}$ and transforms the electromagnetic fields to a new configuration $(\tilde{\mathbf{F}},\tilde{\mathbf{G}},\tilde{\mathbf{J}})$.  
Because the underlying space-time is physically unaltered the manifold is still described by $(M,\mathbf{g},\star)$.  
But for the new field configuration to be physically supported, there must exist a new material distribution $\tilde{\boldsymbol{\chi}}$.  Therefore $\mathrm{d}\tilde{\mathbf{F}}=0$, $\mathrm{d}\tilde{\mathbf{G}}=\tilde{\mathbf{J}}$, and $\tilde{\mathbf{G}}= \tilde{\boldsymbol{\chi}}(\star\tilde{\mathbf{F}})$ holds on $\tilde{M}$.  
Such a transformation could, for example, map $M$ to an image $\tilde{M}$ that contains a hole, i.e.\ a region from which the fields will be excluded, as in the case of an electromagnetic cloak.   

Using the inverse, $\mathcal{T}$, of the map $T$, the initial $\mathbf{F}$ and $\mathbf{G}$ are related to the final $\tilde{\mathbf{F}}$ and $\tilde{\mathbf{G}}$ by the pullback of $\mathcal{T}$, denoted as $\mathcal{T}^*$.  
This implies
\begin{equation} \label{Eq:GPullback}
 \tilde{\mathbf{G}} = \mathcal{T}^*(\mathbf{G}) = \mathcal{T}^*(\boldsymbol{\chi}(\star\mathbf{F})) = \tilde{\boldsymbol{\chi}}(\star\mathcal{T}^*(\mathbf{F})),
\end{equation}
which can be solved for $\tilde{\boldsymbol{\chi}}$ as a function of $x\in\tilde{M}$, giving \cite{Thompson2011jo1,Thompson2011jo2}
\begin{equation} \label{Eq:MaterialChi}
 \tilde{\chi}\indices{_{\lambda\kappa}^{\tau\eta}}(x)= \\ -\Lambda\indices{^{\alpha}_{\lambda}} \Lambda\indices{^{\beta}_{\kappa}} \chi\indices{_{\alpha\beta}^{\mu\nu}}\Big|_{\mathcal{T}(x)} \star\indices{_{\mu\nu}^{\sigma\rho}}\Big|_{\mathcal{T}(x)} (\Lambda^{-1})\indices{^{\pi}_{\sigma}}(\Lambda^{-1})\indices{^{\theta}_{\rho}}\, \star\indices{_{\pi\theta}^{\tau\eta}}.
\end{equation}
In Eq.\ (\ref{Eq:MaterialChi}) $\boldsymbol{\Lambda}$ is the Jacobian matrix of $\mathcal{T}$, $\boldsymbol{\Lambda}^{-1}$ is the matrix inverse of $\boldsymbol{\Lambda}$,  and in solving for $\tilde{\boldsymbol{\chi}}$ we have made use of the fact that on a four-dimensional Lorentzian manifold, acting twice with $\star$ returns the negative, $\star\star\mathbf{F}=-\mathbf{F}$.
The initial material tensor $\boldsymbol{\chi}$ and first $\star$ must be evaluated at $\mathcal{T}(x)$, while everything else is evaluated at $x$.  
Equation (\ref{Eq:MaterialChi}) represents the core of transformation optics in linear dielectric materials. 

Note that $\chi\indices{_{\alpha\beta}^{\mu\nu}}$ on the right hand side of Eq.\ (\ref{Eq:MaterialChi}) need not describe the vacuum, so this approach can be used for transformation optics in arbitrary, non-vacuum, initial linear dielectric media \cite{Thompson2010pra}. 
Furthermore, the space-time is unspecified.  Given a desired space-time, such as near Earth, the appropriate metric enters Eq.\ (\ref{Eq:MaterialChi}) through $\star$, given by Eq.\ (\ref{Eq:star}), as is demonstrated below.

\section{Transformation Optics in Curved Space-Time} \label{Sec:GenRelCorr}
To demonstrate how Eq.\ (\ref{Eq:MaterialChi}) may be applied in a curved space-time, consider a situation where the desired device travels in a stable circular orbit around a spherically symmetric massive body, such as a planet.  
Although in this example the planet is not rotating, it illustrates the idea of a satellite-borne device operating in Earth orbit.  
The space-time around the planet is described by the Schwarzschild metric, which in the Schwarzschild coordinates $(t,r,\theta,\varphi)$ takes the matrix representation
\begin{equation}
 g_{\mu\nu}=\begin{pmatrix} \label{Eq:SchwarzschildMetric}
             -\left(1-\tfrac{2M}{r}\right) & 0 & 0 & 0 \\
             0 & \left(1-\tfrac{2M}{r}\right)^{-1} & 0 & 0\\
             0 & 0 & r^2 & 0\\
             0 & 0 & 0 & r^2\sin^2\theta
            \end{pmatrix},
\end{equation}
where $M$ is the planetary mass.  The contravariant version of the metric, $g^{\mu\nu}$, as needed in Eq.\ (\ref{Eq:star}), is the matrix inverse of $g_{\mu\nu}$.

The utility of transformation optics rests on the specification of a transformation that translates into a device, and one is always free to choose the coordinate system in which to specify a transformation.  
However, certain coordinates lend themselves more readily to particular situations, such as cylindrical coordinates for designing a cylindrical cloak.  
The physical system I have in mind is of light coming in from space.  
The simplest thing to look at is radially ingoing light with frequency defined with respect to Schwarzschild time $t$.  
I want the radially ingoing light to be directed or behave in a particular way, from the Schwarzschild coordinate perspective, that can be specified through a transformation.  Thus it is assumed that the most appropriate coordinates to work in are the Schwarzschild coordinates.

Geometrizing the material parameters into the tensor $\bm{\chi}$ means that $\bm{\chi}$ can be used in a general tensor framework, but interpreting the components of $\bm{\chi}$ as permeability, permittivity, etc. is understood, measured, and built, in a flat, local frame (or Minkowski space-time) according to Eq.\ (\ref{Eq:ConstitutiveComponents2}).  
So while we might want to work in the Schwarzschild coordinates and define the optical transformation there, these coordinates don't make any sense for talking about dielectric materials if you actually wanted to build something, thus we have to express the transformation media in the locally flat frame of an observer orbiting with the device.  
The orbiting observer has his own notion of time with respect to which he is defining spatial hypersurfaces; essentially determining his own breakdown of the field strength tensor and his notion of material.
In such a locally flat frame the metric becomes Minkowskian, and the components of $\bm{\chi}$ may be given an interpretation in terms of the familiar material properties of Eqs.\ (\ref{Eq:ConstitutiveComponents2}).  
An observer co-moving with this frame would measure and interpret the material properties in the same way as in a laboratory on Earth.

\subsection{Construction of a local frame} \label{Sec:Frame}
In flat space-time, one may readily switch between inertial observers by using a Lorentz boost, but switching between inertial observers in a curved space-time is more complicated.  
To express $\bm{\chi}$ in a local frame co-moving with the device, we must first specify the motion of the local frame in Schwarzschild coordinates and then construct a boost-like operation that allows us to switch between the global Schwarzschild coordinates and the local frame.  
Thus, the local frame is essentially the set of Cartesian axes carried by the co-moving observer.

For simplicity, and without great loss of generality, we may restrict attention to motion in the equatorial plane, $\theta = \pi/2$.
The four-velocity of a small object in a circular orbit of radius $r$, in the Schwarzschild coordinates, is \cite{Schutz}
\begin{equation} \label{Eq:FourVelocity}
 u^{\mu} = \frac{1}{\sqrt{1-\tfrac{3M}{r}}}\left(1,0,0,\frac{1}{r}\sqrt{\frac{M}{r}}\right).
\end{equation}
Note that in the Newtonian limit $r \gg M$ this reduces to the usual Newtonian angular velocity
\begin{equation} \label{Eq:AngularVelocity}
 \omega_N = \frac{1}{r}\sqrt{\frac{M}{r}}
\end{equation}
of a particle in circular motion about a mass $M$.
Stable timelike circular orbits only exist for $r\geq 6M$, so the potentially problematic value $r=3M$ is not under consideration.

Let $e_0$ be a proper-time-directed basis vector in the local frame $e_A$.  
The capital Latin indices of the local frame are raised and lowered by the Minkowski metric $\eta_{AB}$.  
In the local frame, the worldline of the co-moving observer is directed along the direction of proper time. 
It follows that in Schwarzschild coordinates, $e_{0}^{\mu}=u^{\mu}$. 
We must now find spatial unit vectors $e_A$ such that $e_A^{\mu}e\indices{_{B\mu}}=\eta_{AB}$.  
Because only $e_{0}^{\mu}$ is constrained, there is some choice as to the orientation of the spatial frame, and the following choice of frame will be discussed further in Sec.\ \ref{Sec:FrameChoice}.  
Let the local observer's $x$-axis coincide with the $r$-direction at all times, which implies
\begin{equation} \label{Eq:Localx}
 e_1^{\mu} = \left(0,\sqrt{1-\frac{2M}{r}},0,0\right).
\end{equation}
A suitable choice for $e_2$ is
\begin{equation} \label{Eq:Localy}
 e_2^{\mu} = \left(0,0,\frac{1}{r},0\right). 
\end{equation}  
Because $e_1$ has only an $r$-component and $e_2$ has only a $\theta$-component, $e_3$ can only have $t$ and $\varphi$ components.  
This, together with the orthogonality conditions, determines
\begin{equation} \label{Eq:Localz}
 e_3^{\mu} = \frac{1}{\sqrt{(r-3M)}}\left(-\sqrt{\frac{Mr}{(r-2M)}},0,0,\frac{\sqrt{r-2M}}{r}\right).
\end{equation}

A transformation matrix for switching between the global Schwarzschild coordinates and the local frame may be constructed with the basis vectors so determined.  A 1-form $n_{\mu}$ in Schwarzschild coordinates is transformed to a 1-form $n_A$ in the local frame by $n_A=S\indices{_A^{\nu}}n_{\nu}$, where
\begin{widetext}
\begin{equation} \label{Eq:frameTransformation}
 S\indices{_A^{\nu}} = \frac{1}{\sqrt{1-\tfrac{3M}{r}}}
 \begin{pmatrix}
  1 & 0 & 0 & \sqrt{\tfrac{M}{r^3}} \\
  0 & \sqrt{\left(1-\tfrac{3M}{r}\right)\left(1-\tfrac{2M}{r}\right)} & 0 & 0\\
  0 & 0 & \tfrac{1}{r}\sqrt{1-\tfrac{3M}{r}} & 0 \\
  -\sqrt{\tfrac{M}{r-2M}} & 0 & 0 & -\tfrac{1}{r}\sqrt{1-\tfrac{2M}{r}}                       
 \end{pmatrix};
\end{equation}
\end{widetext}
while vectors are transformed by $S\indices{^A_{\nu}}$, the transpose of the inverse of $S\indices{_A^{\nu}}$.

One can readily verify that these matrices transform 1-forms and vectors as desired.  
In particular, transforming the Schwarzschild metric to the local frame by
\begin{equation}
 S\indices{_A^{\mu}}S\indices{_B^{\nu}}g_{\mu\nu} = \eta_{AB} =
 \begin{pmatrix}
  -1 & 0 & 0 & 0\\
   0 & 1 & 0 & 0\\
   0 & 0 & 1 & 0\\
   0 & 0 & 0 & 1
 \end{pmatrix},
\end{equation}
returns the Cartesian Minkowski metric, as required (in the equatorial plane).  
I emphasize that in this article we are dealing with two different operations that, perhaps due to an inadequacy of language, are both referred to as ``transformations''.  
In this subsection we have constructed the matrix operation Eq.\ (\ref{Eq:frameTransformation}) that allows us to transform tensors between the global Schwarzschild coordinates and the local co-moving frame of the orbiting device, i.e.\ taking a physical object, such as a vector, expressed in Schwarzschild coordinates and re-expressing it in the local coordinates of the co-moving observer.  
This frame transformation is distinct from the map $\mathcal{T}$ of transformation optics (see Eq.\ (\ref{Eq:GPullback})) that is typically given in terms of a coordinate transformation.  
I will subsequently distinguish these as ``frame transformation'' and ``optical transformation''.

\subsection{Optical Transformations in Schwarzschild Space-Time} \label{Sec:Schwarzschild}
A local, flat, Cartesian frame has been constructed in which the transformation medium parameters have meaning for a co-moving observer, in the sense of Eq.\ (\ref{Eq:ConstitutiveComponents2}).  
As is evident from Eq.\ (\ref{Eq:GPullback}), the optical transformation that must be specified is a map $\mathcal{T}:\tilde{M}\subseteq M \to M$, which is a transformation on the Schwarzschild coordinates of the form
\begin{equation} \label{Eq:GenericTrans}
 \mathcal{T}(t,r,\theta,\varphi) = (t',r',\theta',\varphi') = \\ \left(f_0(t,r,\theta,\varphi),f_1(t,r,\theta,\varphi),f_2(t,r,\theta,\varphi),f_3(t,r,\theta,\varphi)\right).
\end{equation}
Calculating the Jacobian matrix of $\mathcal{T}$ and using Eq.\ (\ref{Eq:MaterialChi}) leads to an expression for $\tilde{\bm{\chi}}$, which may be frame transformed to the local frame by
\begin{equation} \label{Eq:framedChi}
 \hat{\chi}\indices{_{AB}^{CD}}=S\indices{_A^{\mu}}S\indices{_B^{\nu}}S\indices{^{C}_{\sigma}}S\indices{^{D}_{\rho}}\tilde{\chi}\indices{_{\mu\nu}^{\sigma\rho}}.
\end{equation}
The material parameters may be extracted from $\hat{\bm{\chi}}$ and expressed in the representation of Eq.\ (\ref{Eq:ConstitutiveComponents2}).  Note that while $\tilde{\bm{\chi}}$ on the right hand side of Eq.\ (\ref{Eq:framedChi}) is expressed in Schwarzschild coordinates $(t,r,\theta,\varphi)$, $\hat{\bm{\chi}}$ on the left hand side is expressed in the local Cartesian coordinates $(x^0,x^1,x^2,x^3)=(\tau,x,y,z)$.
Generic optical transformations such as Eq.\ (\ref{Eq:GenericTrans}) are computationally intensive, and yield complicated results.  
Consider instead some simpler representative optical transformations.

\subsubsection{Radial Optical Transformations} \label{Sec:RadialTrans}
One application of near Earth transformation optics that springs to mind (disregarding any issues related to near-field or far-field applicability, or boundary conditions) is that of an orbiting telescope fashioned from a diffractionless perfect lens, or ``superlens'' \cite{Pendry2000prl,Grbic2004prl}.  
A perfect lens of width $d$, in the Cartesian coordinates of flat space-time, has been frequently discussed in terms of a linear space-folding optical transformation  \cite{Leonhardt2006njp1,Leonhardt2007jo,Guenneau2010jmo,Lai2011jo} such as
\begin{equation} 
 \mathcal{T}(t,x,y,z)=(t',x',y',z')=
 \begin{cases}
 (t,x,y,z) & x<0, \\
 (t,-ax,y,z) & 0<x<d, a>0, \\
 (t,x-2ad,y,z) & x>d,
 \end{cases}.
\end{equation}


Extrapolating to Schwarzschild space-time, one could imagine a perfect lens in Earth orbit designed to focus radially ingoing light rays, over some small solid angle, with a radius-folding optical transformation.  
However, it has been argued that because the space-folding transformation does not preserve orientation, certain subtleties in the TO procedure mean that this transformation does not adequately describe a superlens \cite{Bergamin2010EMTS}.  
Consider instead the general class of orientation preserving radial optical transformations of the form
\begin{equation} \label{Eq:F1r}
 \mathcal{T}(t,r,\theta,\varphi)=(t',r',\theta',\varphi')=\left(t,f(r),\theta,\varphi\right),
\end{equation}
such that the resulting transformation medium occupies some appropriate radial region and small solid angle.
The Jacobian matrix of the transformation is
\begin{equation}
 \Lambda\indices{^{\alpha}_{\beta}} = \begin{pmatrix}
                                       1 & 0 & 0 & 0\\
                                       0 & f_{,r} & 0 & 0\\
                                       0 & 0 & 1 & 0\\
                                       0 & 0 & 0 & 1
                                      \end{pmatrix},
\end{equation}
and $\star\indices{_{\alpha\beta}^{\mu\nu}}$ is obtained for the Schwarzschild space-time by plugging the Schwarzschild metric, Eq.\ (\ref{Eq:SchwarzschildMetric}), into Eq.\ (\ref{Eq:star}).
The Jacobian matrix $\Lambda\indices{^{\alpha}_{\beta}}$ and $\star\indices{_{\alpha\beta}^{\mu\nu}}$ are now used in Eq.\ (\ref{Eq:MaterialChi}) to calculate the transformation medium $\tilde{\chi}\indices{_{\lambda\kappa}^{\tau\eta}}$.  
Note that because the optical transformation takes place in vacuum, the initial $\chi\indices{_{\alpha\beta}^{\mu\nu}}$, on the right hand side of Eq.\ (\ref{Eq:MaterialChi}), is simply the vacuum $\bm{\chi}$, so that Eq.\ (\ref{Eq:MaterialChi}) reduces to
\begin{equation} \label{Eq:MaterialChiVac}
 \tilde{\chi}\indices{_{\lambda\kappa}^{\tau\eta}}(x)= \\ -\Lambda\indices{^{\alpha}_{\lambda}} \Lambda\indices{^{\beta}_{\kappa}}  \star\indices{_{\alpha\beta}^{\sigma\rho}}\Big|_{\mathcal{T}(x)} (\Lambda^{-1})\indices{^{\pi}_{\sigma}}(\Lambda^{-1})\indices{^{\theta}_{\rho}}\, \star\indices{_{\pi\theta}^{\tau\eta}}.
\end{equation}
Next, this calculated value of $\tilde{\chi}\indices{_{\lambda\kappa}^{\tau\eta}}$ is used together with $S\indices{_A^{\nu}}$ in Eq.\ (\ref{Eq:framedChi}) to find $\hat{\chi}\indices{_{AB}^{CD}}$, which is the material parameter tensor $\bm{\chi}$ expressed in the local, co-moving Cartesian frame.  

Once $\hat{\chi}\indices{_{AB}^{CD}}$ is known, it would be nice to give the material parameters their usual interpretations in terms of permeability, permittivity, etc.  The correspondence between the components of $\hat{\chi}\indices{_{AB}^{CD}}$ and the components of the matrices $\check{\varepsilon}^*$, $\check{\mu}^{-1}$, $\check{\gamma}_1^*$, and $\check{\gamma}_2^*$ in the representation of Eq.\ (\ref{Eq:ConstitutiveComponents1}) are given explicitly in \cite{Thompson2011jo1}, or in the Appendix of \cite{Thompson2011jo2}.  
Once $\check{\varepsilon}^*$, $\check{\mu}^{-1}$, $\check{\gamma}_1^*$, and $\check{\gamma}_2^*$ have been identified, they are then converted, through application of Eqs.\ (\ref{Eq:ConstitutiveShift}), to the usual $\check{\varepsilon}$, $\check{\mu}$, $\check{\gamma}_1$ and $\check{\gamma}_2$ of the representation given by Eq.\ (\ref{Eq:ConstitutiveComponents2}).
These are finally the material parameters that realize the optical transformation Eq.\ (\ref{Eq:F1r}). 
Expressed in the local, co-moving Cartesian frame, they are

\begin{equation} 
\check{\varepsilon} =
  \begin{pmatrix}
   \varepsilon_{xx} & 0 & 0\\
   0 & \varepsilon_{yy} & 0\\
   0 & 0 & \varepsilon_{zz}
  \end{pmatrix} = \frac{fr(r-3M)}{r^3f-M(2r^3+f^3)} 
  \begin{pmatrix}
   \frac{f(f-2M)}{f'(r-2M)} & 0 & 0\\
   0 & rf' & 0 \\
   0 & 0 & \frac{f'(r-2M)\left(r^3f-M(2r^3+f^3)\right)}{r^2(r-3M)(f-2M)}
  \end{pmatrix},
\end{equation}
\begin{equation}
 \check{\gamma}_1 =    \frac{r^3(f-2M)-f^3(r-2M)}{M(2r^2+f^3)-fr^3} \sqrt{\frac{M}{r-2M}}
  \begin{pmatrix}
   0 & 1 & 0\\
   -1 & 0 & 0\\
   0 & 0 & 0
  \end{pmatrix},
\end{equation}
with $\check{\mu}=\check{\varepsilon}$ and $\check{\gamma}_2=\check{\gamma}_2^{\mathtt{T}}$.  This shows that space-time curvature effects appear even for the relatively simple radial transformation of Eq.~(22).  In this case it is straightforward to see that in the limit $r\gg M$ and for localized transformations $f(r)\sim r$ this reduces to the Newtonian result
\begin{equation}
 \check{\varepsilon} = 
 \begin{pmatrix}
  1/f' & 0 & 0 \\
  0 & f' & 0\\
  0 & 0 & f'
 \end{pmatrix}, \quad \check{\gamma}_1=0.
\end{equation}

Typically, dielectric motion requires magnetoelectric coupling terms and also tends to change the required values of $\varepsilon$ and $\mu$ \cite{Landau8,Thompson2011jo1}, indeed this motional effect has been used to interpret magnetoelectric couplings as velocities in transformation optics \cite{Leonhardt2006njp1}. 
But this does not happen for the optical transformation Eq.\ (\ref{Eq:F1r}), even though the device is in relative motion throughout its orbit.  
Why not? 
Consider for a moment electromagnetic fields measured in a laboratory in flat space-time.  
If a dielectric medium moves through the lab with speed $v$, then the fields experienced by the medium are not those measured in the lab frame, but are instead the lab-measured fields boosted to the medium frame. 
By boosting to the medium frame, one finds that medium traveling perpendicularly to the transformation direction requires no modification over stationary medium, while medium moving parallel to the transformation direction requires velocity dependent modifications to the material parameters of the stationary medium.
Returning now to the orbiting device, we have chosen an orbiting frame such that the instantaneous velocity is always perpendicular to the $r$-direction, the direction of the action of the optical transformation.  Therefore, in this particular case we should not expect any contribution from the velocity. 

Mass dependent, general relativistic contributions start to appear even for the relatively simple transformation Eq.~(22).  As a slightly more general class of radial transformations, consider
\begin{equation} \label{Eq:F1rtheta}
 \mathcal{T}(t,r,\theta,\varphi)=(t',r',\theta',\varphi')=\left(t,g(r,\varphi),\theta,\varphi\right)
\end{equation}
such that the equivalent material occupies some appropriate radial region and small solid angle.
In this case the corresponding material parameters, in the local Cartesian frame and occupying the appropriate spatial region, are
\begin{equation} \label{Eq:F1rthetamu}
 \check{\mu}= \check{\varepsilon} = \frac{g(g-2M)(r-3M)}{A}
 \begin{pmatrix}
 \frac{r\left(2Mg-g^2-g_{,\varphi}^2\right)}{(2M-r) g_{,r}} 
 & 0 
 & \frac{r^{3/2} g_{,\varphi}}{\sqrt{r-3M}} \\
 0 
 & r^2  g_{,r}
 & 0 \\
 \frac{r^{3/2} g_{,\varphi}}{\sqrt{r-3M}}   
 & 0 
 & \frac{(2 M-r) \left[M(2r^3+g^3)-gr^3\right] g_{,r}}{(2M-g)(3M-r)r}
 \end{pmatrix}
\end{equation}

\begin{equation}\label{Eq:F1rthetagamma}
 \check{\gamma}_{1} = \check{\gamma}_2^{\mathtt{T}} = \frac{\sqrt{M(r-2M)}}{A} 
 \begin{pmatrix}
  0 & H  & 0 \\
  -H  & 0 & -\sqrt{r(r-3M)}g^2g_{,r}g_{,\varphi} \\
  0 & \sqrt{r(r-3M)}g^2g_{,r}g_{,\varphi} & 0
 \end{pmatrix}
\end{equation}
where everything is evaluated on the equator,
\begin{equation}
H = gr(r^2-g^2)-\frac{g-2M}{r-2M}\left(2M(r^3-g^3) \right) -g^2g_{,\varphi}^2,
\end{equation}
and
\begin{equation}
 A = (g-2M) \left[r^3g-M(2r^3+g^3)\right]+M g^2 g_{,\varphi}^2.
\end{equation}
%
where everything is evaluated on the equator, and
\begin{equation}
 A = r(r-2M)(r-3M)-Mg_{,\varphi}.
\end{equation}
Due to the presence of $M$ in Eqs. (\ref{Eq:F1rthetamu}) and (\ref{Eq:F1rthetagamma}) it is clear that there is some additional contribution over what might be expected from a similar optical transformation leading to a stationary device in flat space-time.  
However, distinguishing the contribution due to the orbital velocity from that of the space-time curvature requires some additional effort.
Before turning to such an analysis, consider an optical transformation that mixes space and time components.

\subsubsection{Optical Transformations Mixing Space and Time} \label{Sec:SpaceTimeTrans}
Recently, optical transformations mixing space and time have generated some interest \cite{Leonhardt2006njp1,Cummer2011jo,Thompson2011jo2}, and have been examined in the context of frequency converting active metamaterials \cite{Cummer2011jo}.  
The transformation discussed in Ref.\ \cite{Cummer2011jo} was given by
\begin{equation} \label{Eq:xdepTime}
 \mathcal{T}(t,x,y,z)=(t',x',y',z')=\left(\frac{t}{ax+b},x,y,z\right)
\end{equation}
within a stationary slab in flat space-time. For our purposes, consider a similar optical transformation, in the Schwarzschild coordinates,
\begin{equation} \label{Eq:F0tr}
 \mathcal{T}(t,r,\theta,\varphi)=(t',r',\theta',\varphi')=\left(f_0(t,r),r,\theta,\varphi\right),
\end{equation}  
piecewise defined over a suitable radial region and small solid angle.  
It is assumed that the portions of the optical transformation outside this small active region is boundary matched and leaves the external vacuum region unaffected, but may not be smooth at the boundary.
In this case the corresponding material parameters, in the local, co-moving, Cartesian frame, are found to be
\begin{widetext}
 \begin{equation} \label{Eq:F0trResult1}
  \check{\mu}=\check{\varepsilon} = \frac{1}{B}
   \begin{pmatrix}
    (3M-r)f_{0,t} & 0 & \left(1-\tfrac{2M}{r}\right)\sqrt{M(r-3M)}f_{0,r} \\
    0 & (3M-r)f_{0,t} & 0\\
    \left(1-\tfrac{2M}{r}\right)\sqrt{M(r-3M)}f_{0,r} & 0 & f_{0,t}^{-1}\left[B-M\left(1-\tfrac{2M}{r}\right)^2f_{0,r}^2\right]
   \end{pmatrix},
 \end{equation}
 \begin{equation} \label{Eq:F0trResult2}
  \check{\gamma}_1=\check{\gamma}_2^{\mathtt{T}} = \frac{\sqrt{r-2M}}{B}
   \begin{pmatrix}
    0 & -\sqrt{M}(1-f_{0,t}^2) & 0 \\
    \sqrt{M}(1-f_{0,t}^2) & 0 & -\left(1-\tfrac{2M}{r}\right)\sqrt{r-3M}f_{0,t}f_{0,r} \\
    0 & \left(1-\tfrac{2M}{r}\right)\sqrt{r-3M}f_{0,t}f_{0,r} & 0
   \end{pmatrix},
 \end{equation}
\end{widetext}
where
\begin{equation}
 B=M-(r-2M)f_{0,t}^2.
\end{equation}
This demonstrates another example where a class of optical transformations in an orbit of Schwarzschild space-time requires an $M$-dependent transformation medium.  
Again the question remains of how to disentangle the velocity and curvature contributions, since both of these will appear as $M$ dependencies. 
More precisely, how can one determine the curvature induced contribution over what one would obtain for a device performing the same action when in orbit around a Newtonian planet of mass $M$ residing in a flat, Minkowskian background?  
We will return to this question in Sec.\ \ref{Sec:Discussion}, but first consider one more example of interest.

\subsubsection{Orbiting Cloaking Device} \label{Sec:OrbitingCloakTrans}
\begin{figure}[ht]
 \scalebox{1}{\includegraphics{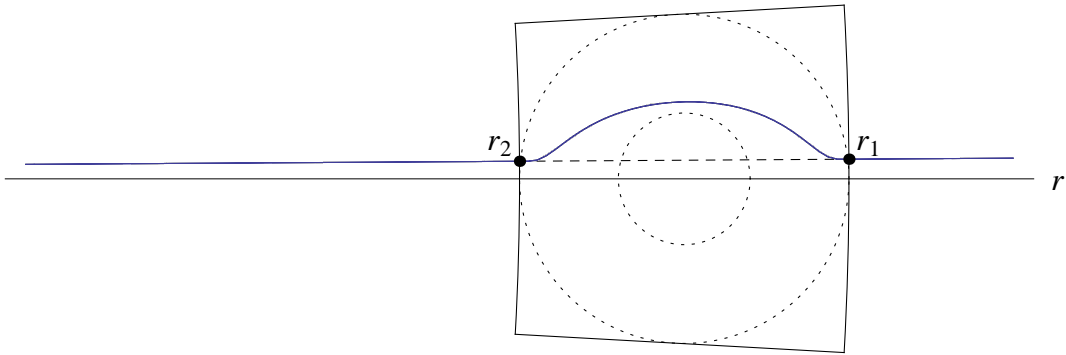}}
 \caption{A radially ingoing ray enters the cloak at the point $r_1=r_c+r_o$ and exits at the point $r_2=r_c-r_o$.  Instead of following the chord connecting $r_1$ and $r_2$ (dashed line), light is deflected upon entering the cloak and follows a diverted path (solid line). The diverted path is described by a bump function and is therefore $C^{\infty}$ at the cloak boundary.}
 \label{Fig:ChordBump}
\end{figure}

Suppose it is desired to construct an orbiting cloaking device that hides a region from radially ingoing light rays.  
Let the device occupy a segment of an annulus that circumscribes a sphere of radius $r_o$, and let the interior region be of a size to contain a sphere of radius $r_i$, as in Fig.\ \ref{Fig:ChordBump}.
Let coordinates on the vacuum manifold be $(t',r',\theta',\varphi')$ and coordinates on the manifold containing the cloak be $(t,r,\theta,\varphi)$.  
For simplicity we may consider a two-dimensional cross-section of the cloak lying in the equatorial plane $\theta=\pi/2$.
The three-dimensional cloak can obtained by either by rotating the two-dimensional cross-section about the radial line passing through its center, or, for a different cloak geometry, rotating the two-dimensional cross section through an azimuthal angle $\Delta\theta$, such that the inscribed circles of Fig.\ \ref{Fig:ChordBump} become arced cylinders.
Let the center of the cloak orbit at a radius $r_c$, and be instantaneously located at $\varphi=0$. 
A radially ingoing ray enters the cloak at a point $r_1 = r_c+r_o$ and exits the cloak at point $r_2=r_c-r_o$, as in Fig.\ \ref{Fig:ChordBump}. 

\begin{figure}[ht]
 \scalebox{1}{\includegraphics{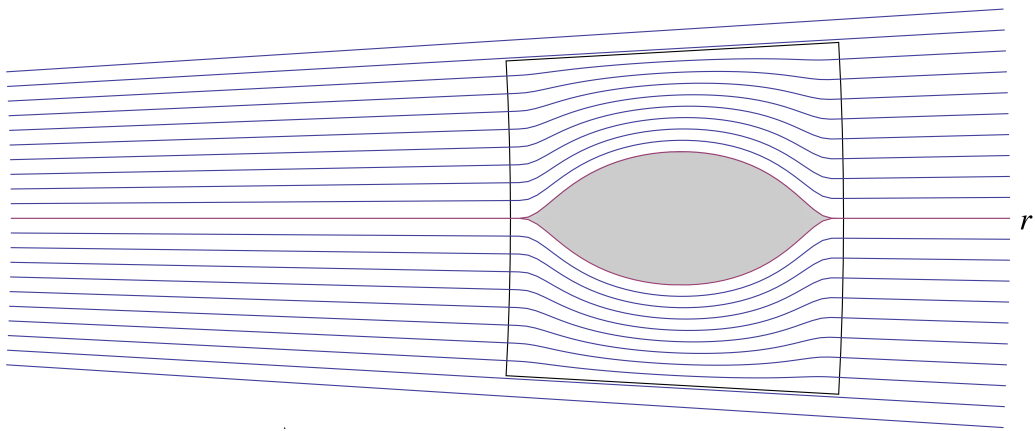}}
 \caption{Radially ingoing light rays follow a deflected trajectory as they pass through the cloak. The cloaked region (shaded) is not spherical, but is designed to contain a sphere of radius $r_i$ and angular size $2\omega_i$.  The deflection amplitude varies with the angle of the ingoing ray and has maximum value $\omega_i$ for $\varphi=0$ and decreases linearly to zero at $\varphi=\omega_o$.}
 \label{Fig:AnnularBump}
\end{figure}

Instead of following the chord connecting $r_1$ and $r_2$, the ray is deflected upon entering the cloak and follows a diverted path.  
Such a path may be described via a map $T:M\to\tilde{M}$ from the chord in the vacuum manifold to the manifold containing the cloak.  This may be thought of as a coordinate transformation that rotates a point $(r',\varphi')$, lying on the chord, to a point $(r=r',\varphi(r',\varphi'))$ on the diverted path.
The maximum amount of deflection from a given chord will depend on the ingoing angle $\varphi'$, as in Fig.\ \ref{Fig:AnnularBump}.  
An ingoing ray with $\varphi'=0$ has a maximum deflection equal to the half-angle spanned by the inner spherical region (e.g.\ small dotted circle in Fig.\ \ref{Fig:ChordBump})
\begin{equation}
 \omega_i = \sin^{-1}\left(\frac{r_i}{r_c}\right).
\end{equation}
An ingoing ray with $\varphi'=\omega_o$, where
\begin{equation} 
 \omega_o = \sin^{-1}\left(\frac{r_o}{r_c}\right)
\end{equation}
is the half-angular size of the cloak (or the sphere circumscribed by the cloak - e.g.\ large dotted circle in Fig.\ \ref{Fig:ChordBump}),
grazes the exterior surface of the cloak and will not be deflected at all.

To find a map satisfying these conditions, let $\varphi$ be given by
\begin{equation} \label{Eq:ChordBump}
 \varphi(r'=r,\varphi') = \varphi' + \left(\omega_i-\frac{\omega_i}{\omega_o}\varphi'\right)b(r),
\end{equation}
where 
\begin{equation} \label{Eq:BumpFunction}
 b(r) = \begin{cases}
         \exp\left[1-\frac{r_o^2}{r_o^2-(r-r_c)^2}\right] & (r_c-r_o) < r < (r_c+r_o), \\
         0 & \text{elsewhere}
        \end{cases}
\end{equation}
is a ``bump function'' of unit amplitude.  Although a bump function is piecewise defined, it is $C^{\infty}$ everywhere and therefore smooth across the boundary. For calculational purposes assume $\varphi'\geq 0$, allowing the cloak parameters to be specified in the upper part of Fig.\ \ref{Fig:AnnularBump}. The lower part may be obtained by reflection.

Recall from Sec.\ \ref{Sec:TO} that, while it may be convenient to describe the desired trajectory with a map $T:M\to\tilde{M}$, such as Eq.\ (\ref{Eq:ChordBump}), from the vacuum manifold to the manifold with material, it is the pullback of the inverse map $\mathcal{T}:\tilde{M}\to M$ that maps the  electromagnetic field strength and excitation tensors $\bm{\mathrm{F}}$ and $\bm{\mathrm{G}}$ from the vacuum manifold to the manifold with material. It is only then that the corresponding trajectory of radially ingoing light rays passing through the cloak material will be diverted as in Fig.\ \ref{Fig:AnnularBump}.  Therefore, let
\begin{equation} \label{Eq:OrbitingCloakTrans}
 \mathcal{T}(t,r,\theta,\varphi) = (t',r',\theta',\varphi') = \left(t,r,\theta,f_3(r,\varphi)\right)
\end{equation}
where
\begin{equation}
 f_3(r,\varphi)=\frac{\varphi-\omega_ib(r)}{1-\frac{\omega_i}{\omega_o}b(r)}, \quad \varphi \geq \omega_ib(r).
\end{equation}

From Eqs.\ (\ref{Eq:MaterialChi}), (\ref{Eq:frameTransformation}) and (\ref{Eq:framedChi}), the material parameters for the cloak, expressed in the local Cartesian frame and representation of Eq.\ (\ref{Eq:ConstitutiveComponents2}), are
\begin{widetext}
 \begin{equation} \label{Eq:CloakMu}
  \check{\mu}=\check{\varepsilon} = \frac{1}{C}
   \begin{pmatrix}
    (3M-r)f_{3,\varphi} & 0 & (2M-r)\sqrt{r(r-3M)}f_{3,r} \\
    0 & (3M-r)f_{3,\varphi} & 0\\
    \left(2M-r\right)\sqrt{r(r-3M)}f_{3,r} & 0 & f_{3,\varphi}^{-1}\left[C-r\left(2M-r\right)^2f_{3,r}^2\right]
   \end{pmatrix},
 \end{equation}
 \begin{equation} \label{Eq:CloakGamma}
  \check{\gamma}_2=\check{\gamma}_1^{\mathtt{T}} = \frac{\sqrt{M(r-2M)}}{C}
   \begin{pmatrix}
    0 & -(1-f_{3,\varphi}^2) & 0 \\
    1-f_{3,\varphi}^2 & 0 & -\sqrt{r(r-3M)}f_{3,\varphi}f_{3,r} \\
    0 & \sqrt{r(r-3M)}f_{3,\varphi}f_{3,r} & 0
   \end{pmatrix},
 \end{equation}
\end{widetext}
where 
\begin{equation}
 C = 2M - r + Mf_{3,\varphi}^2.
\end{equation}

\begin{figure}[ht]
\subfigure[$\ \epsilon_{xx}=\epsilon_{yy}$]{
   \includegraphics[scale=.4] {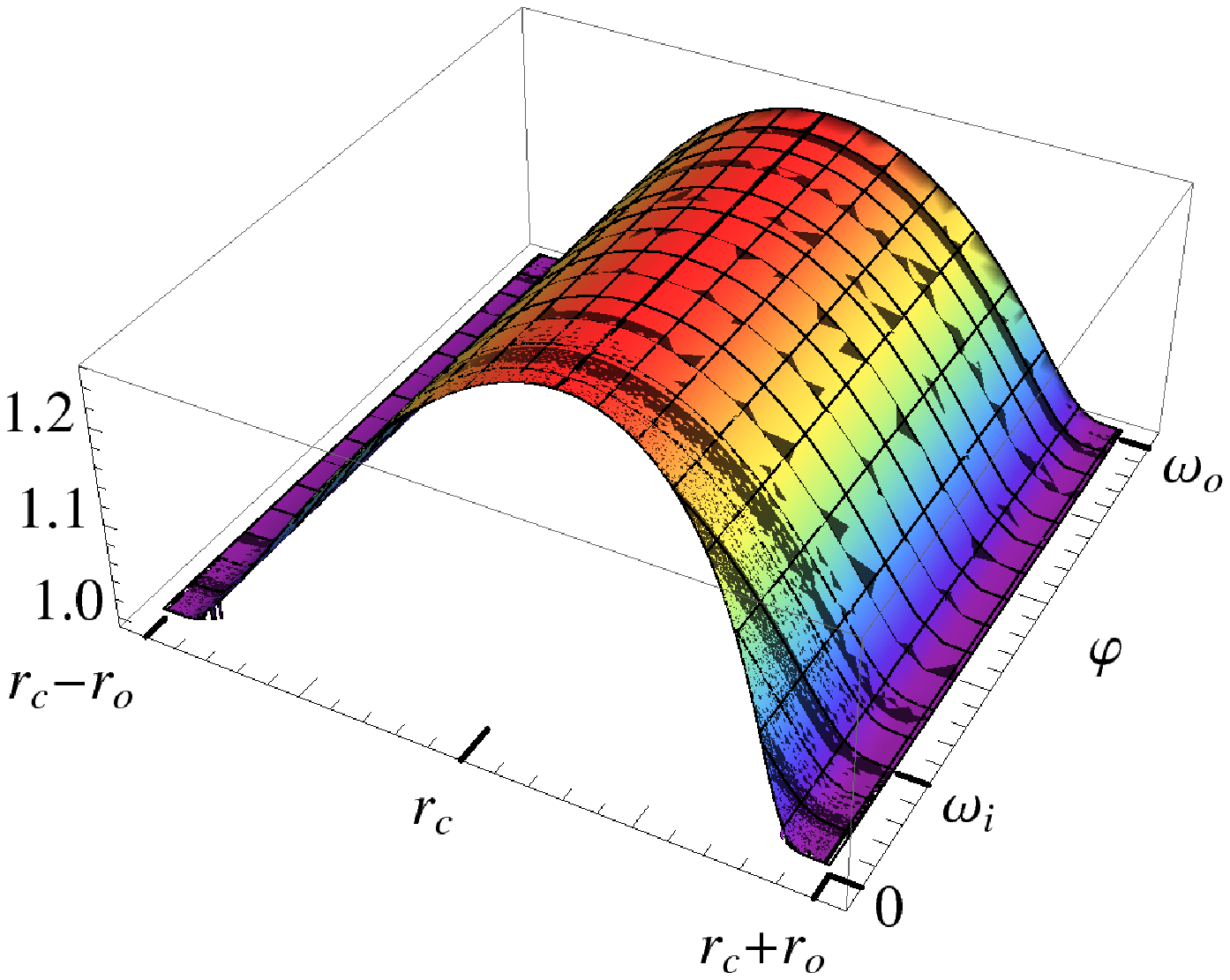}
   \label{Fig:epsilon_xx}
 }
 \subfigure[$\ \epsilon_{xz}$]{
   \includegraphics[scale=.4] {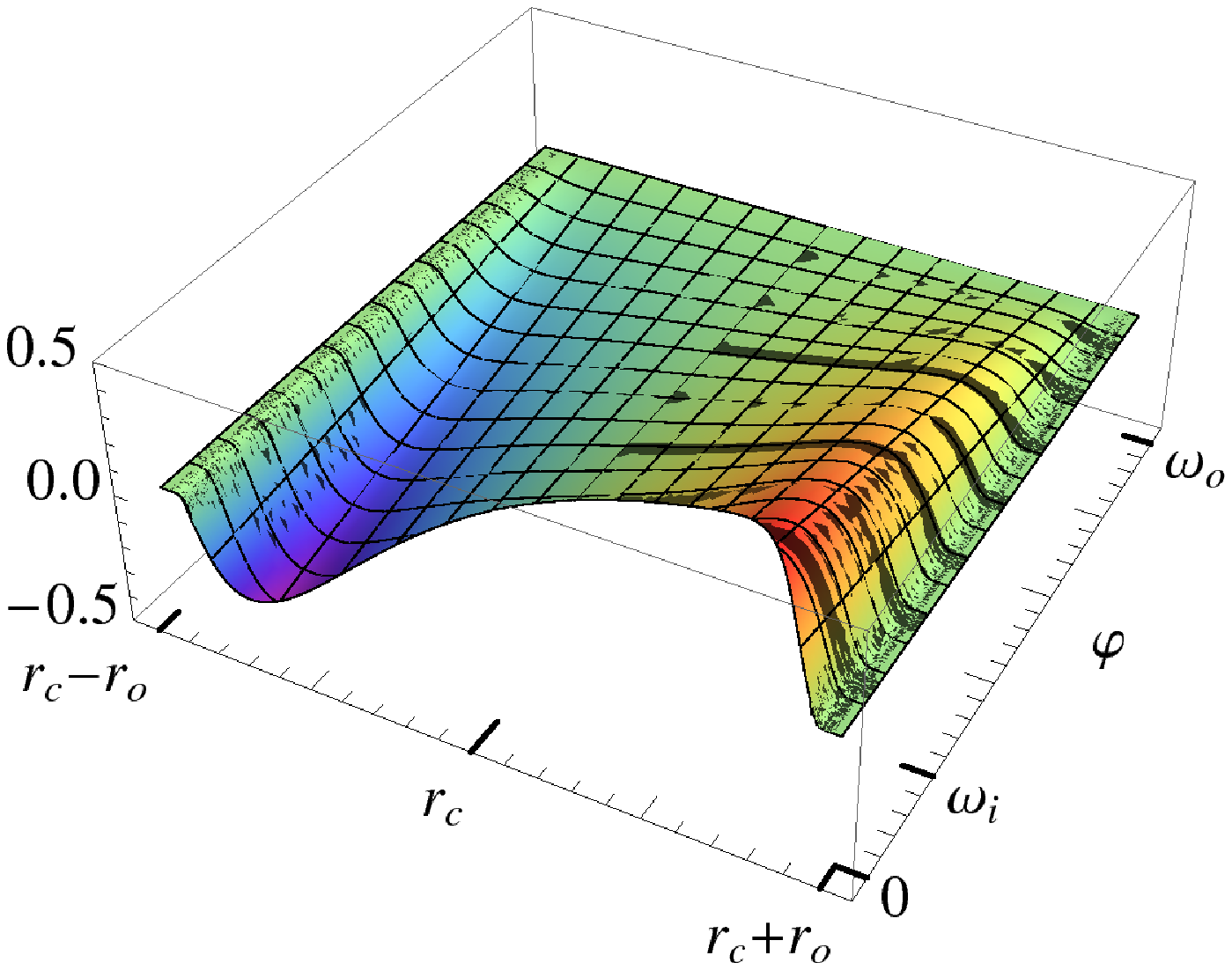}
   \label{Fig:epsilon_xz}
 }
 \subfigure[$\ \epsilon_{zz}$]{
   \includegraphics[scale=.4] {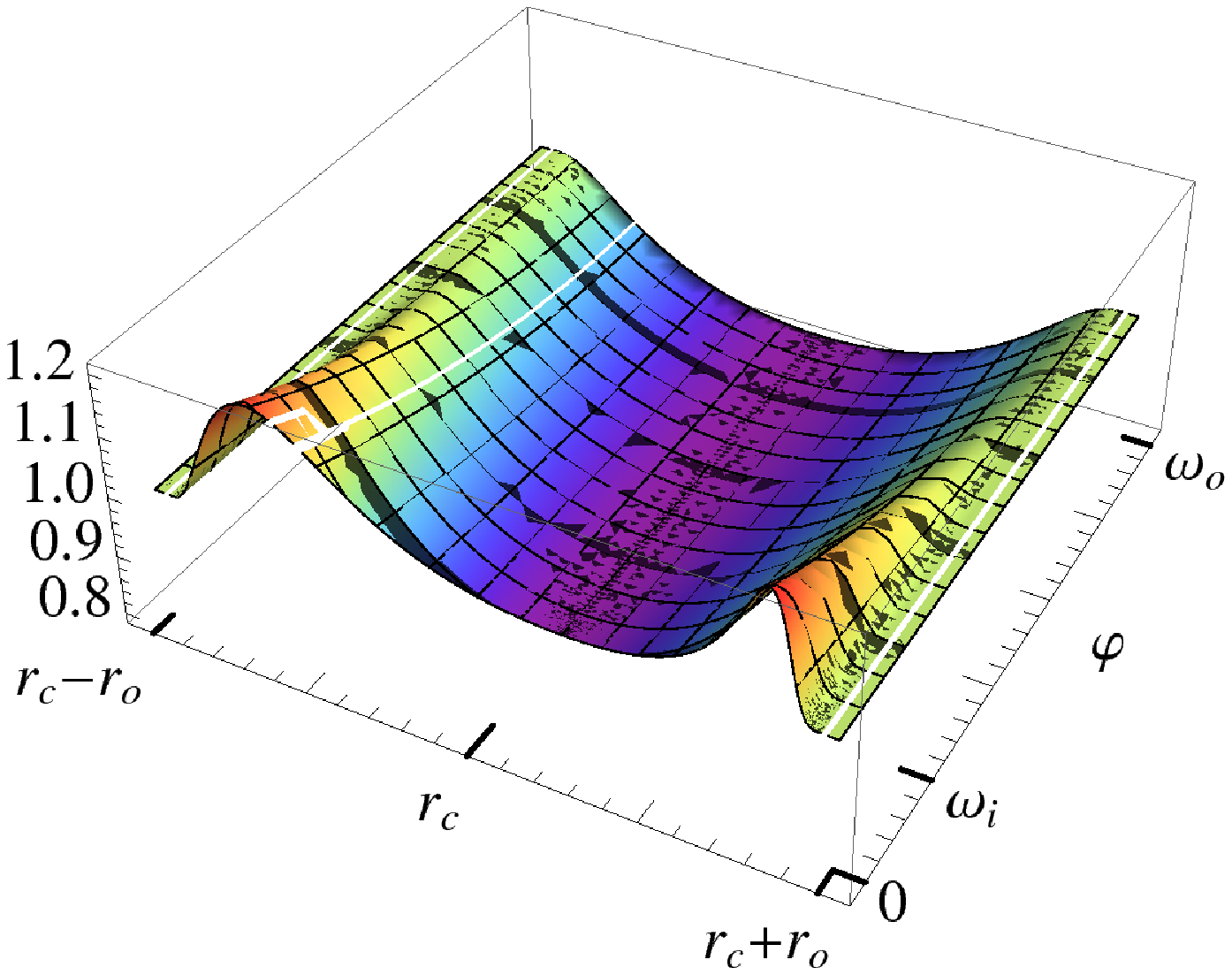}
   \label{Fig:epsilon_zz}
 }

\subfigure[$\ \gamma^2_{xy}$]{
   \includegraphics[scale=.4] {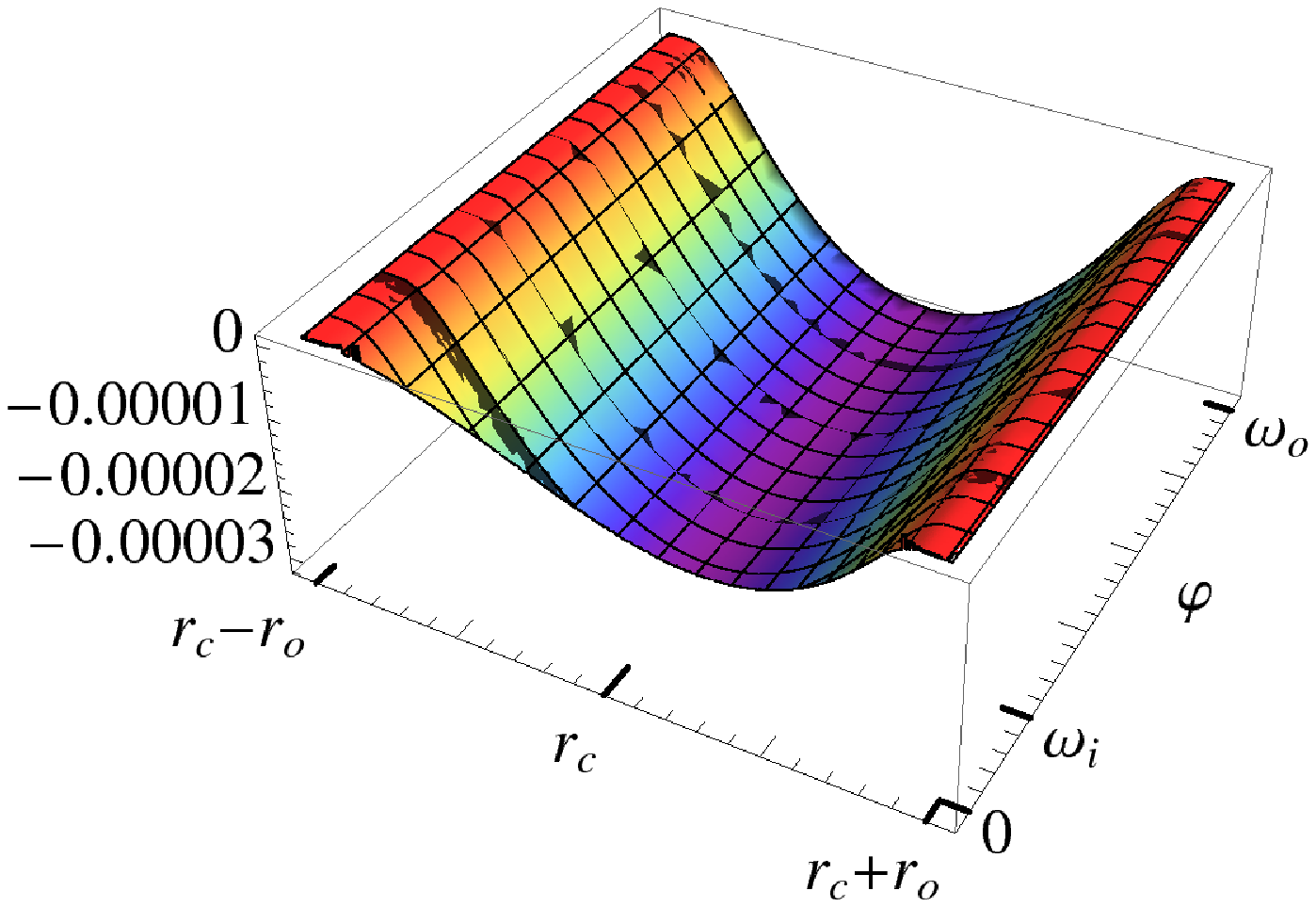}
   \label{Fig:gamma2_xy}
 }
 \subfigure[$\ \gamma^2_{zy}$]{
   \includegraphics[scale=.4] {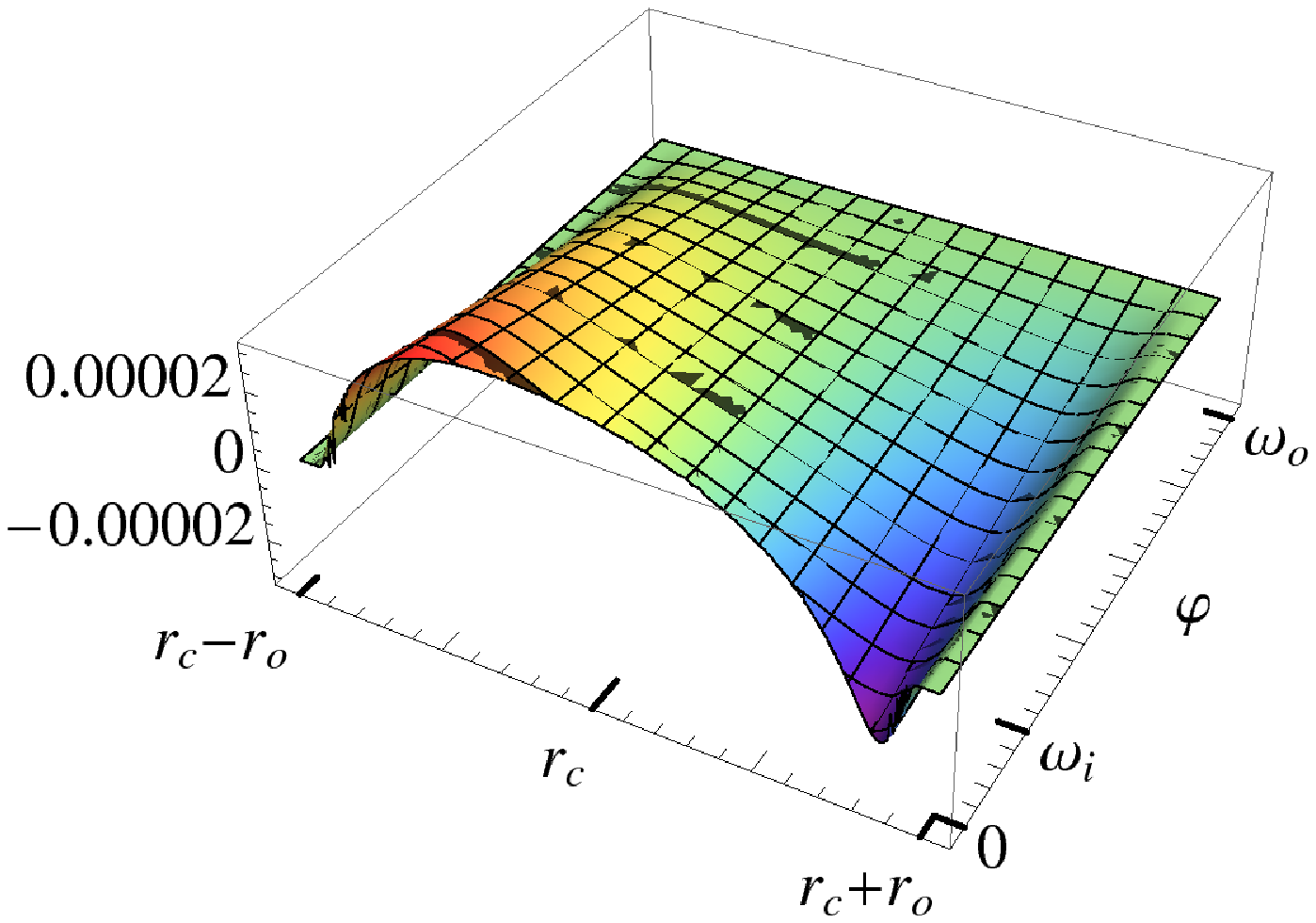}
   \label{Fig:gamma2_zy}
 }
\caption{Material parameters for half of a two-dimensional slice of an orbiting cloaking device described by the transformation given in Eq.\ (\ref{Eq:OrbitingCloakTrans}). A three-dimensional cloak can be constructed in two ways: either by rotating about the $r$ axis, or by reflection across the $r$ axis and then stacking copies of the slice to form a tube-like hidden region.}
 \label{Fig:Cloak_Parameters}
\end{figure}

Consider a specific example of such a cloak.  The orbital radius of a typical satellite is $r_c\sim 10^8 cm$, and the geometrized mass of Earth is $M_E\approx 0.3 cm$.  Let $r_i=100 cm$ and $r_o=500 cm$.  The corresponding material parameters for such a cloak are plotted in Fig. \ref{Fig:Cloak_Parameters}.

Because the angular size of the cloak is so small, the curved shape, greatly exaggerated in Figs.\ \ref{Fig:ChordBump} and \ref{Fig:AnnularBump}, is not perceptible.  The density plots of Fig.\ \ref{Fig:Cloak_Parameters} look very much as they would for a similarly defined cloak in flat space, which returns us to the question posed at the end of the Sec.\ \ref{Sec:SpaceTimeTrans}, namely, how can one determine the curvature induced contribution to the material over what one would obtain for a device performing the same action in a flat, Minkowskian background?  This is addressed in the next section.

\section{Discussion} \label{Sec:Discussion}
It was mentioned earlier that for transformation optics, a dielectric moving with respect to the frame in which the optical transformation has been defined typically requires magnetoelectric couplings, and that in cases where an optical transformation results in such couplings, they can sometimes be interpreted as a velocity \cite{Leonhardt2006njp1,Thompson2011jo1}.
Because the calculations of Sec.\ \ref{Sec:GenRelCorr} have been demonstrated for a material in orbital motion, and because the four-velocity is expressed in terms of the planetary mass and orbital radius, it is justified to ask whether the $M$-dependent results truly include space-time curvature effects or are merely a consequence of the dielectric's orbital motion, after all, the Riemann curvature tensor does not appear explicitly in any of the calculations.  But the Riemann tensor need not appear explicitly to find effects that only appear in curved space-times.

The ideal approach would be to compare with results obtained in the absence of general relativity, i.e.\ the transformation media required for the same optical transformations applied in conjunction with a Newtonian theory of gravity.
For this purpose, consider a ``Newtonian'' system consisting of the dielectric device in circular orbit around a spherical mass $M$, where the background space-time is Minkowskian, i.e. a system with vanishing space-time curvature.  
This assumption combines the relativistic covariance of electrodynamics with Newtonian gravity, and essentially consists of a dielectric device moving in a circle of radius $r$ in Minkowski space-time, with angular velocity given by Eq.\ (\ref{Eq:AngularVelocity}).  We may still consider optical transformations in spherical coordinates, and must still express the dielectric material properties in the local frame of the device.

At first glance it might appear that since the correct Newtonian velocity is obtained by taking the Newtonian limit $r\gg M$ of Eq.\ (\ref{Eq:FourVelocity}), one should proceed by using the Newtonian limit of Eq.\ (\ref{Eq:frameTransformation}) to perform the same calculations.  
This approach is incorrect.  
In flat space-time, frame transformations between inertial observers is accomplished with a Lorentz boost, but the Newtonian limit of of Eq.\ (\ref{Eq:frameTransformation}) is not a Lorentz boost. 
A circular orbit in flat space-time is non-inertial, and for such a motion there is no clearly accepted method for frame transforming between observers over the entire orbital path.  
The best we can do is use a Lorentz boost to frame transform to an instantaneously co-moving frame.

\subsection{Comparison with Boosted Frames} \label{Sec:BoostCompare}
The first thing we can do is compare the fully relativistic results with those expected in a boosted frame in flat space-time, which will reveal whether or not the full results arise solely from the orbital motion.
Since it is assumed that the transformations in the Schwarzschild coordinates occur over a very small solid angle, and the final results are expressed in a locally flat Cartesian frame, we may compare with an equivalent Cartesian transformation in Minkowski space-time. 
For definiteness, return to the transformation of Eq.\ (\ref{Eq:F0tr}).  Due to the orientation of the local Cartesian frame chosen in Eqs.\ (\ref{Eq:Localx}) -- (\ref{Eq:Localz}), this corresponds to a transformation of the form
\begin{equation} \label{Eq:ComparisonTrans}
  \mathcal{T}(t,x,y,z)=(t',x',y',z')=\left(f_0(t,x),x,y,z\right),
\end{equation}
boosted in the $z$ direction with some speed $\beta$. A Lorentz boost in the $z$ direction has matrix representation \cite{MTW}
\begin{equation}
 L\indices{^{\mu}_{a}} = 
  \begin{pmatrix}
   \gamma & 0 & 0 & \gamma\beta\\
   0 & 1 & 0 & 0\\
   0 & 0 & 1 & 0\\
   \gamma\beta & 0 & 0 & \gamma,
  \end{pmatrix}
\end{equation}
where $\gamma = (1-\beta^2)^{-1/2}$.  
This frame-transforms components of a 1-form (or lower tensor indices), $\omega_a=\omega_{\mu}L\indices{^{\mu}_{a}}$, such that the frame with Latin indices moves with velocity $\beta \hat{z}$ in the frame with Greek indices.  Components of a vector (or upper tensor indices) are frame-transformed by $v^a=L\indices{^{a}_{\mu}}v^{\mu}$, where $L\indices{^{a}_{\mu}}$ is obtained by setting $\beta\to -\beta$ in $L\indices{^{\mu}_{a}}$.

Equation (\ref{Eq:MaterialChi}) is perfectly compatible with such boosts \cite{Thompson2011jo2}. Applying Eq.\ (\ref{Eq:MaterialChi}) and a boost to the transformation of Eq.\ (\ref{Eq:ComparisonTrans}) in Minkowski space-time results in the material parameters
\begin{widetext}
 \begin{equation} \label{Eq:ComparisonResult1}
  \check{\mu}=\check{\varepsilon} = \frac{1}{\beta^2-f_{0,t}^2}
  \begin{pmatrix}
   (\beta^2-1)f_{0,t} & 0 & \beta\sqrt{1-\beta^2}f_{0,r} \\
   0 & (\beta^2-1)f_{0,t} &  0 \\
   \beta\sqrt{1-\beta^2}f_{0,r} & 0 & f_{0,t}^{-1}\left(\beta^2-\beta^2f_{0,r}^2-f_{0,t}^2\right)
  \end{pmatrix},
 \end{equation}
 \begin{equation} \label{Eq:ComparisonResult2}
  \check{\gamma}_1=\check{\gamma}_2^{\mathtt{T}} = \frac{f_{0,t}}{\beta^2-f_{0,t}^2}
  \begin{pmatrix}
   0 & -\beta(1-f_{0,t}) & 0 \\
   \beta(1-f_{0,t}) & 0 & -\sqrt{1-\beta^2}f_{0,r} \\
   0 & \sqrt{1-\beta^2}f_{0,r} & 0
  \end{pmatrix}.
 \end{equation}
\end{widetext}
By comparing Eqs.\ (\ref{Eq:ComparisonResult1}) and (\ref{Eq:ComparisonResult2}) with Eqs.\ (\ref{Eq:F0trResult1}) and (\ref{Eq:F0trResult2}), it is clear that the fully relativistic results Eqs.\ (\ref{Eq:F0trResult1}) and (\ref{Eq:F0trResult2}) cannot arise from a pure boost in flat space-time.
Choosing 
\begin{equation} \label{Eq:EquivVelocity}
 \beta = \pm\sqrt{M/(r-2M)}
\end{equation}
will recover the desired values of $\check{\mu}_{xx}$, $\check{\mu}_{yy}$, and $\check{\gamma}_{1xy}$ (for the negative root of $\beta$) but not the other non-zero components.  
Choosing instead
\begin{equation}
 \beta = \pm \frac{(r-2M)\sqrt{M}f_{0,t}}{\sqrt{4M^2(M-r)+r^2(r-2M)f_{0,t}^2}}
\end{equation}
will recover, if real, the desired value of $\check{\mu}_{zz}$ but none of the other non-zero components.
However, there is no boost speed that will recover the desired $\check{\mu}_{xz}$ or $\check{\gamma}_{1yz}$ components.  This demonstrates that the fully relativistic results obtained in Eqs.\ (\ref{Eq:F0trResult1}) and (\ref{Eq:F0trResult2}) are not equivalent to a boost in flat Minkowski space-time, and are therefore not solely a consequence of the orbital motion -- there are space-time curvature contributions.

\subsection{Newtonian Limit} \label{Sec:NewtonianLimit}
Section \ref{Sec:BoostCompare} demonstrates that the fully relativistic results are, in general, not simply a result of the orbital motion but also include curvature effects.  
But the curvature contribution was not isolated.  
If scientists of the future wanted to send a transformation optics designed probe to a black hole, general relativistic effects would surely become important.  
What about for near Earth applications?  
In the geometrized units used here, mass may be measured in distance.  The geometrized mass of Earth is roughly $0.3cm$, while the radius of a typical satellite orbit is on the order of $10^8cm$.  
Additionally, the velocity of a satellite in orbit is much less than the speed of light, so $\beta<<1$.  Thus the next thing we can do is look at an expansion of the results in the Newtonian limit.

From Eq.\ (\ref{Eq:EquivVelocity}) it should be expected that in the non-relativistic, Newtonian limits $\beta\to 0$ and $r>>M$ (equivalently $M\to 0$), Eqs.\ (\ref{Eq:F0trResult1}) and (\ref{Eq:F0trResult2}) are equivalent to the flat space-time results boosted with the Newtonian speed for circular orbits, $\beta=\sqrt{M/r}$.  
Expanding Eqs.\ (\ref{Eq:F0trResult1}) and (\ref{Eq:F0trResult2}) to lowest order in $M$, and Eqs.\ (\ref{Eq:ComparisonResult1}) and (\ref{Eq:ComparisonResult2}) to second order in $\beta$ shows this is indeed the case.  
Continuing with the expansion, one finds curvature terms appearing in the next order of the expansion, in terms proportional to 
\begin{equation} \label{Eq:Corrections}
 \frac{M}{r}\sqrt{\frac{M}{r}}.
\end{equation}
Therefore, it is found that the requisite material parameters, described in a local, co-moving, Cartesian frame can be described as
\begin{equation}
 \varepsilon = \varepsilon_0 + \varepsilon_1 + \varepsilon_2 +O\left(\frac{M^2}{r^2}\right)
\end{equation}
where $\varepsilon_0$ are the material parameters for an equivalent transformation in flat Minkowski space-time, $\varepsilon_1$ are terms proportional $M/r$, and $\varepsilon_2$ are terms proportional to Eq.\ (\ref{Eq:Corrections}).
While only the velocity contributes to $\varepsilon_1$, both velocity and curvature contribute terms in $\varepsilon_2$.
 
Near the horizon of a black hole, these corrections can approach $\varepsilon_1\approx 0.5, \varepsilon_2\approx .35$, which can represent a significant contribution for transformation media with $\varepsilon\sim 1$.  On the other hand, for a satellite orbiting Earth with $r\sim 10^8$cm these corrections are on the order of $\varepsilon_1\sim 10^{-9}$ and $\varepsilon_2\sim 10^{-13}$.  With light, differences of 1 part in $10^{13}$ are not difficult to measure with an interferometer, but such tiny corrections are not likely to have great significance for most applications of transformation optics in Earth orbit. On the other hand, in gradient indexed materials comprising holes or voids in glass the void volume scales with the index of refraction, thus one would have to control the void volume to these levels.  Such precision is certainly challenging but is not inconceivable, thus it may be that such corrections will be required for highly sensitive applications in the future.  

\subsection{Frame Choice} \label{Sec:FrameChoice}
The particular frame described in Sec.\ \ref{Sec:Frame}, is not the only choice of local frame.  In fact, since the local x-axis is always radially aligned, the chosen frame is actually spinning with respect to the distant stars.  
It may be of interest to instead consider a non-spinning local frame.  This kind of frame is said to be ``Fermi-Walker'' transported \cite{MTW,Klein2008cqg}.  
Since in the Fermi-Walker case the x-axis will not always be pointed radially outward, the parametrization of the local frame will be time dependent and  
the required frame transformations will also be much more complicated. 
The spinning frame of Sec.\ \ref{Sec:Frame} was chosen to simplify the demonstration of general relativistic corrections.

It was shown in Sec.\ \ref{Sec:BoostCompare} that the fully relativistic results are not, in general, a consequence of the translational dielectric motion alone, but could they simply be a gyroscopic effect of the spinning frame?  
Indeed, the Lorentz transformations consist of both boosts and rotations, and the fact that pure boosts do not form a closed subgroup of the Lorentz group gives rise to, for example, Thomas Precession \cite{Thomas1926nat,Jackson}.
However, it is easy to see that Eqs.\ (\ref{Eq:F0trResult1}) and (\ref{Eq:F0trResult2}) cannot be obtained as a combined boost and rotation in flat space-time.  Such a result would be obtained in a similar manner to Eqs.\ (\ref{Eq:ComparisonResult1}) and (\ref{Eq:ComparisonResult2}), although in this case the Lorentz transformation is slightly more complicated.

A general Lorentz transformation is of the form
\begin{equation}
 \bm{L}=\exp[\bm{\omega}\cdot\bm{S}+\bm{\xi}\cdot\bm{K}]
\end{equation}
where the $4\times 4$ matrices $S_i$ and $K_i$ are the generators of rotations and boosts, $\bm{\omega}$ is a 3-vector denoting an angle about an axis of rotation, and
\begin{equation}
 \bm{\xi}=\tanh^{-1}(\beta)\, \bm{\hat{\beta}}
\end{equation}
is the boost 3-vector  \cite{Jackson}.  
In this case we may restrict ourselves to a boost in the $z$ direction and a rotation in the $xz$ plane.  
Taking the non-relativistic, Newtonian limits shows that the limiting behaviour of Eqs.\ (\ref{Eq:F0trResult1}) and (\ref{Eq:F0trResult2}) cannot be recovered for any non-zero $\bm{\omega}$.  Thus the full result is not simply an artifact of boosts and rotations, but does in fact include curvature contributions.

\subsection{Geometric Optics} \label{Sec:GeometricOptics}

While I have calculated the general relativistic, curvature induced contributions to the material parameters, one might also like to compare the difference in the propagation of light through the corrected and uncorrected materials by solving Maxwell's equations through the medium, e.g.\ by ray tracing through the cloak of Sec.\ \ref{Sec:OrbitingCloakTrans}.  
Of course, for the full material parameters of Eqs.\ (\ref{Eq:CloakMu}) and (\ref{Eq:CloakGamma}), the rays follow the trajectories of Fig.\ \ref{Fig:ChordBump} \textit{by design}, but the question for a TO designed instrument would be whether the uncorrected parameters are ``good enough'' for the desired application.
This means it would be desirable to compare the exact result with the ray trajectories through a device constructed from the material parameters obtained in the Newtonian limit, but still orbiting in the Schwarzschild geometry. 
Since GR contributions are generally weak, the material parameters and boundaries generally complicated, and the space-time curved, the calculation would likely require a more complicated dislocation approach \cite{KlineKay} rather than an eikonal or plane-wave ansatz in order to retain sufficient sensitivity in the geometric optics limit.
Such a calculation lies outside the scope of the current article, the purpose of which is as the first foray of transformation optics in curved space-times; I therefore leave this for further study.

\section{Conclusions} \label{Sec:Conclusions}
The completely covariant approach to transformation optics introduced in Refs.\ \cite{Thompson2011jo1,Thompson2011jo2} allows for arbitrary background space-times.  
This enables space-time curvature effects to be accounted for in transformation optics.  
I have demonstrated the particular case of optical transformations for transformation media in a stable circular orbit of the spherically symmetric Schwarzschild geometry, simulating an Earth-orbiting satellite.
In order to have their usual physical interpretations, the material parameters must be expressed in a locally flat Cartesian frame, co-moving with the satellite.  
After constructing such a frame, and the matrices used to frame transform $\bm{\chi}$ into it, some generalized optical transformations were considered to illustrate the calculations.  
As a specific example, a unidirectional orbiting cloak that redirects radially ingoing rays around a central cavity was considered in Sec.\ \ref{Sec:OrbitingCloakTrans}.

Of particular concern is whether the mass contributions to the results are a consequence of the space-time curvature or the orbital velocity, as both depend on  $M$.  
However, it was explicitly demonstrated for the optical transformation given by Eq.\ (\ref{Eq:F0tr}) that the result is not equivalent to motion without curvature, i.e.\ motion in flat space-time.
The curvature effects were further disentangled by considering the non-relativistic, Newtonian limits.  In these limits, the results are in fact obtainable as a pure boost in flat space-time, with no rotation.  Higher order terms of the expansion provide curvature induced corrections.

The transformation considered in detail, Eq.\ (\ref{Eq:F0tr}), generalizes a transformation that has been previously studied in the context of frequency shifting metamaterials.  
However, a similar frequency shifting effect, called the gravitational redshift, arises naturally from the gravitational potential \cite{MTW} and is easily measurable on Earth \cite{Pound1960prl}.  
It is, therefore, of no surprise that when overlaying a frequency altering transformation on the naturally occurring gravitational redshift, space-time curvature effects will be required to end up with the correct, desired frequency.  
Such corrections are small, but are potentially meaningful and accessible for future high precision applications of transformation optics.

\acknowledgments
I thank J\"org Frauendiener and Steve Cummer for useful comments and discussions, John Hannay for useful comments on the manuscript, and Gerrard Liddell for assistance with the plots.


%

\end{document}